
\input harvmac
\noblackbox
\def\lieg{{\underline{\bf g}}}
\def\liet{{\underline{\bf t}}}
\def\Tr{\rm Tr}
\def\p{\partial}
\def\pl{Phys. Lett.}
\font\cmss=cmss10 \font\cmsss=cmss10 at 7pt

\def\IR{\relax{\rm I\kern-.18em R}}
\def\IC{\relax\hbox{$\inbar\kern-.3em{\rm C}$}}
\def\inbar{\,\vrule height1.5ex width.4pt depth0pt}

\def\IZ{\relax\ifmmode\mathchoice
{\hbox{\cmss Z\kern-.4em Z}}{\hbox{\cmss Z\kern-.4em Z}}
{\lower.9pt\hbox{\cmsss Z\kern-.4em Z}}
{\lower1.2pt\hbox{\cmsss Z\kern-.4em Z}}\else{\cmss Z\kern-.4em
Z}\fi}

\lref\withoo{G. 't Hooft, in {\sl Recent Developments in Gauge Theories},
ed. G. 't Hooft {\it et al. }, Plenum Press (1980);
E. Witten, Nucl. Phys. {\bf B202} (1982) 253.}
\lref\vw{C. Vafa  and E. Witten, Nucl. Phys. {\bf B431}  (1994) 3,
hep-th/9408074; see also
L. Girardello, A. Giveon, M. Porrati
and A. Zaffaroni, Phys.Lett. {\bf B334} (1994) 331, hep-th/9406128.}
\lref\gno{P. Goddard, J. Nuyts, and D. Olive,
Nucl. Phys. {\bf B125} (1977)1; P. Goddard
and D. Olive,
Nucl. Phys. {\bf B191} (1981)528. }
\lref\bourbaki{N. Bourbaki, {\it Groupes et Algebres
de Lie} , ch. 6, Masson, 1981.}
\lref\humphreys{J.E. Humphreys,  {\it Introduction to
Lie Algebras and Representation Theory} Springer 1972. }.
\lref\mol{C. Montonen and D. Olive, Phys. Lett. {\bf 72B} (1977) 117.}
\lref\nisf{H. Osborne,
Phys. Lett.
{\bf 83B}(1979)321.}
\lref\sen{A. Sen, Int.J.Mod.Phys. {\bf A9}  (1994) 3707,  hep-th/9402002 \semi
Phys. Lett. {\bf B329} (1994) 217, hep-th/9402032.}
\lref\wol{E. Witten and D. Olive, Phys. Lett. {\bf 78B} (1978) 97.}
\lref\crdy{J. Cardy and E. Rabinovici, Nucl. Phys. {\bf B205} (1982) 1;
J. Cardy, Nucl. Phys. {\bf B205} (1982) 17.}
\lref\shap{A. Shapere and F. Wilczek, Nucl. Phys. {\bf B320} (1989) 669. }
\lref\filq{A. Font, L. Ibanez, D. Lust, and F. Quevedo, Phys. Lett. {\bf 249B}
(1990) 35;S. J. Rey, Phys. Rev. {\bf D43} (1991) 526;
A. Strominger, Nucl. Phys. 343B (1990) 167; C. Callan, J. Harvey
 and A. Strominger, Nucl. Phys. 359B (1991) 40;
 M. Duff, Class. Quant. Grav. 5 (1988) 189; M. Duff and J. Lu,
 Nucl. Phys. 354B (1991) 129; 357B (1991) 354; Phys. Rev. Lett. 66 (1991)
 1402; Class. Quant. Grav. 9 (1991) 1; M. Duff, R. Khuri and J. Lu, Nucl.
 Phys. 377B (1992) 281; J. Dixon, M. Duff and J. Plefka, preprint
 CTP-TAMU-60/92, hep-th/9208055; I. Pesando and A. Tollsten, Phys. Lett. 274B
 (1992) 374;J. Harvey, J. Gauntlett and J. Liu, Nucl. Phys. 409B (1993)
 363; R. Khuri, \pl 259B (1991) 261; J. Gauntlett and J. Harvey,
 hep-th/9407111; J.H. Schwarz and A. Sen, Nucl. Phys. 404B (1993) 109.}
\lref\olivrev{D. Olive,  in {\it Monopoles in quantum field theory: }
proceedings
of the Monopole Meeting, Trieste Italy, December 1981 edited by N. S. Craigie,
P. Goddard
and W. Nahm, Singapore World Scientific 1982.}
\Title{ \vbox{\baselineskip12pt\hbox{hep-th/9501022}
\hbox{EFI-95-01 }
\hbox{YCTP-P2-95}}}{\vbox{
\centerline{Reducing $S$-Duality  to $T$-Duality }}}
\centerline{{ Jeffrey A. Harvey}}
\vskip.05in
\centerline{\sl Enrico Fermi Institute}
\centerline{\sl University of Chicago}
\centerline{\sl 5640 Ellis Avenue, Chicago, IL 60637}
\vskip .1in
\centerline{ Gregory Moore}
\vskip.05in
\centerline{\sl Department of Physics}
\centerline{\sl Yale University}
\centerline{\sl New Haven, CT 06520}
\vskip.1in
\centerline{\sl and}
\vskip.05in
\centerline{ Andrew Strominger}
\vskip.1in
\centerline{\sl Department of Physics}
\centerline{\sl University of California}
\centerline{\sl Santa Barbara, CA 93106-9530}
\vskip.15in
\noindent ABSTRACT:
The infrared limit of $D=4,~~N=4$
Yang-Mills
theory with compact
gauge group $G$ compactified
 on a two-torus is governed by
an effective superconformal
field theory. We conjecture that this is a certain orbifold
involving the maximal torus of $G$.
Yang-Mills  $S$-duality makes predictions for
all correlators of this effective conformal
field theory. These predictions are shown to be implied by the
standard $T$-duality of the
conformal field theory. Consequently,
Montonen-Olive duality between electric
and magnetic states reduces to the standard
two-dimensional duality between momentum and
winding states.

\Date{Jan. 8, 1995; Revised Feb. 23, 1995}

Seventeen years ago Montonen and Olive \mol\
made a bold conjecture:
Yang-Mills theory with gauge group $G$
and coupling constant $e$ is {\it identical}
to a gauge theory  based on the dual
gauge group $G^v$ \foot{We recall the definition
of $G^v$ below.} \gno\
with  coupling $4\pi /e$. The identification
involves a relabeling of states and operators,
interchanging particles with  solitons, and electric
charges with magnetic charges. It was
quickly realized \refs{\wol,\nisf} that the conjecture is viable only for $N=4$
supersymmetric Yang-Mills. When a $\theta$ angle is included, $2 \pi$
shifts of $\theta$ together with the
$\IZ_2$ symmetry of Montonen and Olive generate
an $SL(2,\IZ)$ symmetry  which acts on the complex coupling constant
\eqn\tgt{ \tau \equiv{\theta \over 2 \pi}+{4 \pi  \over e ^2} i
\equiv {\theta \over 2 \pi}+{i \over \alpha}
}
as
\eqn\stu{\tau \rightarrow {a \tau +b \over c \tau +d},}
with $ad-bc=1,~~~a,b,c,d\in \IZ$.
This $SL(2,\IZ)$ symmetry on the space of theories is known as
$S$-duality.  $S$-duality originated in the study of
lattice models \refs{\crdy, \shap}  but has come to play
a prominent role in recent speculations concerning the structure
of both $N=4$ Yang-Mills theory and string theory \refs{\filq }
as reviewed in \sen.

Initial evidence in favor of $S$-duality for $N=4$ Yang-Mills
was provided by the exact  agreement \refs{\wol,\nisf} of the
(calculable) masses of the stable elementary particles and solitons
with those predicted by the $\IZ_2$ symmetry of Montonen and Olive.
More recently  the
masses of  some bound states have been demonstrated to be in agreement
with the full $S$-duality \sen, and the predictions of $S$-duality
for a topologically twisted version of the theory on
a more general four-manifold have been tested \vw\ .

A skeptic could remain unconvinced by this evidence.  It
concerns only zero-momentum, supersymmetric or topological properties of the
theory.
Such properties are highly constrained by the powerful
symmetries of theory, especially by the
$N=4$ supersymmetry. Thus, a true skeptic
may argue that the evidence to date all follows from the known symmetries of
the theory in some delicate way.  A more reasonable skeptic might argue that
$S$-duality is indeed non-trivial, but only holds for the supersymmetric or
BPS saturated states of the theory.
If the  Montonen-Olive
conjecture is correct, the theory and its $S$-dual must agree on
much more than this. In particular,  all finite-momentum
correlation functions must agree.
This is clearly {\it not} implied by the known
symmetries:
the addition of higher dimension operators to the theory can change the
correlation functions without affecting the topological quantities.
It is also clear that a two-particle state with non-zero center of mass
momentum is not BPS saturated even if the individual one-particle
states are. Thus evidence for $S$-duality at non-zero momentum
necessarily involves evidence for $S$-duality away from the supersymmetric
subspace of the theory.

In this paper,  we will
propose and confirm -- with some assumptions -- a
finite-momentum test of  $S$-duality, albeit in a  very special limit.
The idea is to compactify four-dimensional $N=4$ Yang-Mills with
group $G$
to two dimensions on a torus. At distances large
compared to the size of the
torus, the effective theory must reduce to a conformal
field theory. We conjecture and give plausibility arguments
that this takes a particular form
involving the maximal torus $T\subset G$ (and
some antisymmetric tensor fields if $\theta$ is nonzero).
$S$-duality transformations involve no dimensionful parameters,
and therefore commute with scale transformations. $S$-duality of
$D=4,~~N=4$ Yang-Mills therefore makes a definite
prediction of an exact duality symmetry of the effective conformal field
theory  which must act on all finite-momentum correlation
functions. This prediction will indeed be confirmed in the following:
$S$-duality
reduces to the well-known ``$T$-duality'' of conformal field
theory in which tori are interchanged with their duals. The interchange of
electric and magnetic charges effected by $S$-duality is essentially  the
familiar  interchange of momentum and winding modes
effected by $T$-duality.

Consider  compactification of
$D=4, N=4$ supersymmetric Yang-Mills
theory with compact gauge group $G$. For
simplicity we first consider only the case
for which the simply connected covering group $\tilde G$
associated to $G$ is $SU(n)$. The general case is treated in
the appendix.

The bosonic part of the $D=4,~~N=4$
action is
\eqn\ii{
\eqalign{
I^{bosonic} &= -{1\over 4 \pi} {1\over  \alpha}
\int d^4x \sqrt{-g^{(4)}} \Tr \Biggl[
 \half F_{\mu\nu} F^{\mu \nu} + \sum_{I=1}^6 D_\mu \phi^ID^\mu \phi^I
+  \sum_{1\leq I<J\leq 6} ([\phi^I,\phi^J])^2 \Biggr] \cr
& -   {\theta \over 8 \pi^2} \int \Tr F\wedge F \quad , \cr}
}
where all fields take values in the Lie algebra
$\lieg$ of $G$, and $\Tr $ is a nondegenerate
bilinear form on $\lieg$.
We normalize $\Tr$ so that Euclidean instantons with
integral winding number $k$ have action $2\pi i k \tau$.  If
$\lieg={\bf su(n)} $ is identified with the Lie algebra of $n\times n$
antihermitian matrices the metric is
\foot{The normalization for arbitrary
simple $G$ is given in
eq. $(A.1)$. }
\eqn\metlie{
(a,b)_{{\bf su(n)} }  \equiv - Tr_{\IC^n} (ab).
}

We now compactify the theory by
taking spacetime to be $\Sigma\times T^2$ where
the internal space $T^2$ is  a small two torus of volume $ L^2$.
The line element is
\eqn\llml{ds^2 = - (d \sigma^0)^2 + (d \sigma^1)^2 +
{L^2 \over \rho_2} \mid dx^2+ \rho dx^3\mid^2,}
where $0<x^2,x^3 \le 1$, and
$\rho=\rho_1+i \rho_2$ is a modular parameter
for the internal torus.
Define  fields $X^I\in\lieg$, $I=1,8$ by
\eqn\flds{
\eqalign{
X^I & \equiv  L \phi^I \qquad I = 1,6\cr
X^7 & \equiv  A_2 \cr
X^8 & \equiv  A_3 \quad . \cr}
}
The effective action at length scales
much greater than $L$ then reduces to
\eqn\vi{
\eqalign{
I^{bosonic} = -{1\over 4 \pi} {1\over  \alpha}
&
\int_{\Sigma} d^2\sigma  \Tr \Biggl[
 {L^2 \over 2} F^2 +
 \sum_{1\leq I,J\leq 8} G_{IJ} D_\mu X^I D^\mu X^J\cr
&
+ {1 \over L^2} \sum_{I<J} [X^I,X^J][X^K,X^L]G_{IK} G_{JL}
 \Biggr] \cr
-{\theta \over 4 \pi^2}
\int_{\Sigma}
&
\Tr\Biggl[F_{01} [X^7,X^8] - D_0 X^7 D_1 X^8 +
 D_1 X^7 D_0 X^8\Biggr]  d\sigma^0 \wedge d\sigma^1  \cr}
}
where $\mu,\nu = 0,1$, and the metric $G_{IJ}$ is
\eqn\metric{
G_{IJ} = \delta^{(6)}_{IJ} \oplus {1\over \rho_2}
\pmatrix{\mid \rho\mid^2 & - \rho_1\cr -\rho_1 & 1\cr}\quad .
}
There are in addition fermionic terms
whose form is fixed by the extended
supersymmetry  but these will not be needed.

The action \vi\ contains terms of dimension not equal to
two and so does not represent the infrared limit of the
theory. Our conjecture is that in the infrared limit
it flows to a conformal theory  with bosonic action:
\eqn\xiii{
\eqalign{
I^{bosonic} = -{1\over 4 \pi} {1\over  \alpha}&
\int d^2\sigma \Tr \Biggl[ \sum_{1\leq I,J\leq 8} G_{IJ} \p_\mu X^I \p^\mu
X^J\Biggr]
\cr
-{\theta \over 4 \pi^2} &
\int \Tr\Biggl[ - \partial_0 X^7 \partial_1 X^8 +
 \partial_1 X^7 \partial_0 X^8\Biggr]  d\sigma^0 \wedge d\sigma^1.
\cr}
}
In contrast to \vi, the fields
$X^I$, $I=1,8$,  now take values in the
Cartan subalgebra $\liet$. Moreover, they are
subject to important global identifications discussed below.

A very naive argument leading to \xiii\ is the following. The third
term in \vi\ is a potential term for the
 $X$'s. It is relevant and grows in the
infrared. At low energies $X$ is thus restricted to values
for which the potential vanishes, namely the
Cartan subalgebra $\liet$. For $X$ in $\liet$, the charge current of the
$X$'s vanish. The gauge fields
(which have no local dynamics in two dimensions) may then be completely
decoupled from the $X$'s in light cone gauge.  Their action is
quadratic and they may be integrated out.

This argument is too naive for several reasons. Consider\foot{We
are grateful to D. Kutasov, E. Martinec and N. Nekrasov
for discussion on these
matters.}
an $N=2$ Landau-Ginzburg model with a $\phi^n$
potential. The potential
is relevant and for
$n=2$ one indeed finds that the infrared limit is the (trivial) theory
with $\phi$ restricted to the minimum of the potential, in accord with
the preceding paragraph. However for $n>2$ there is no mass gap and
the infrared limit is a minimal model determined by $n$.
So in general it is not correct  simply to restrict the
fields to the
minimum of the potential.
For our model, if we
denote by $Y$ fields orthogonal to the Cartan subalgebra
and by $Z$ fields in the Cartan subalgebra,  then in
the infrared limit $L \rightarrow 0$
\vi\ contains large quartic interactions of the form $Y^4$ and $Z^2 Y^2$.
In four spacetime dimensions these terms give a mass to the $Y$ fields
at generic points in the moduli space of
$D=4,N=4$ vacua, and the IR limit is just a theory of
the $Z's$ (that is, an abelian gauge theory).  The action \xiii\ is just the
dimensional reduction of this abelian theory.
Unfortunately, the compactification along
$\Sigma\times T^2$ is not so simple  because  the
large wavelength fluctuations of the $Z$ fields explore all of the moduli
space.
Near $Z \sim 0$ the $Y$ fields are light and must be taken into account.
As $L \rightarrow 0$ the region in $Z$ field space
for which the $Y$ fields
are light becomes vanishingly small. Thus in the infrared limit we might be
able to restrict attention to the $Z$ fields with the $Y $ fields set to zero,
and
the two-dimensional gauge fields decoupled.  If the $Y$ fields do appear
in some way in the infrared limit then the two-dimensional gauge fields
do not decouple and must be dealt with.
Clearly a more careful analysis, perhaps using the
$N=4$ supersymmetric non-renormalization theorems,
 is required to see if the above assumptions are justified.
Despite these misgivings, we strongly suspect that \xiii\ is
at least part of the answer.  We henceforth proceed on that assumption.

We now discuss global identifications of the fields $X^I\in \liet$
appearing in \xiii.
Two types of identifications
arise because choosing the Cartan subalgebra $\liet$  does
not completely fix the gauge freedom.
\foot{Related observations were
made long ago in \withoo.}
 First of all,
we must take into account gauge transformations
of the form
\eqn\gts{
g(x_2,x_3) = \exp\biggl[ 2 \pi x_2 A + 2 \pi x_3 B
\biggr]
}
where $A,B\in \liet$ must satisfy
\eqn\crooti{
\exp 2 \pi A = \exp 2 \pi B =1\qquad
}
in order that \gts\ is single valued as a nontrivial loop is traversed in the
internal torus. The set of such Lie algebra elements
 forms the coweight lattice
$\Lambda_{coweight}(G)\subset \liet$:
\eqn\dfcroot{
  \Lambda_{coweight}(G)\equiv
\{ A\in \liet: \exp (2 \pi A ) =1 \}\subset \liet
}
$\Lambda_{coweight}(G)$ is the dual to the weight lattice $\Lambda_{weight}(G)$
of $G$.
A gauge transformation of the form \gts\ shifts $A_2$ and $A_3$ by
$A$ and $B$.  We must therefore identify:
\eqn\xxix{
X^{7}\sim  X^{7} + 2 \pi v
}
\eqn\xxixb{
X^{8}\sim  X^{8} + 2 \pi v '
}
where $ v,  v' \in \Lambda_{coweight}(G)$ are coweight vectors.
There are no such identifications of $X^1, ... X^6$ since they
descend from $D=4$ scalars rather than gauge fields.  Secondly,
we must
identify all fields by the action of $W(G)$, the Weyl group
of $G$.
We conclude that the proper domain for
$(X^1, \dots, X^6; X^7,X^8)$ is the orbifold:
\eqn\xiv{
\Biggl\{ \liet^6 \times \bigl[\liet /2 \pi \Lambda_{coweight}(G)
\bigr] \times \bigl[ \liet /2 \pi \Lambda_{coweight}(G)
\bigr]  \Biggr\}/ W(G)
}
where $W(G)$ is the Weyl group, acting diagonally
on all $X^I$.

Finally, there are $8\times {\rm  rank}(G)$, real,
left and right-moving fermions. The dimensional
reduction of a $D=4,N=4$ theory yields a
$D=2,N=4$ theory\foot{The target space \xiv\  is naturally
a hyperk\"ahler manifold since it can be written
as
$$
\Biggl[ \liet^4 \times T^* \biggl\{ (\liet\otimes \IC)/
(2 \pi[ \Lambda_{coweight}(G) + i  \Lambda_{coweight}(G) \bigr])
\biggr\} \Biggr]/W(G)
$$
},  so
 we arrive at the result:
{\it The leading $L\to 0$ behavior of the
D=4, N=4 SYM theory is governed by the
$D=2$,
$\hat {c}=8\times {\rm  rank}(G)$,
$(4,4)$ superconformal
field theory of an
orbifold with target space defined by \xiv. }

To be quite explicit,
take $\theta=\rho_1=0, \rho_2=1$, and
use
the metric on $\lieg=\bf{su(n)}$ given in \metlie\  and an
orthonormal basis $T^j$ to
identify $\liet$ with $\IR^{n-1}$ and the metric
\metlie\ with Euclidean metric:
\eqn\lieeuc{
\eqalign{
\liet \cong \IR^{n-1} : \sum_{j=1}^{n-1} x_j T^j & \rightarrow
(x_1, \dots, x_{n-1}) \cr
(X^{I=1,6}, X^7 , X^8) &
\rightarrow (\vec X^{I=1,6}, \vec X^7 , \vec X^8) \quad \vec X^J
 \in \IR^{n-1} \quad J=1,8\quad . \cr}
}
The action for the conformal field theory is then given by:
\eqn\cfti{
I^{bosonic} = {1\over 4 \pi \alpha} \int d^2 \sigma
\sum_{J=1}^8
\biggl[   \p_0 \vec X_j \cdot \p_0 \vec X^J
-
\p_1 \vec X^J \cdot \p_1 \vec X^J
\biggr]
}
where $\vec X^J \in \IR^{n-1}$ is identified by
\eqn\cftii{
\eqalign{
\vec X^{7} & \sim \vec X^{7} + 2 \pi  \vec u\cr
\vec X^{8} & \sim \vec X^{8} + 2 \pi  \vec u'
\qquad \vec u,\vec u'
 \in   \Lambda_{coweight} (G) \cr
(\vec X^1,\dots , \vec X^8)  & \sim
 ( w\cdot \vec X^1, \dots,  w\cdot \vec X^8) \qquad w\in W(G) \cr}
}

$S$-duality exchanges a gauge group with
its dual $G^v$,  the magnetic
group of Goddard, Nuyts,
and Olive (GNO) \gno \olivrev.
\foot{$G^v$ is also known as
 the ``Langlands dual'' in the mathematics
literature.}
The global
structure of a simple compact Lie group $G$ is
given by specifying either its  weight lattice or coweight
lattice.  GNO noticed that the coweight lattice of $G$
is always the {\it weight} lattice of a dual group $G^v$.
In the case where $G$ and $G^v$ have the same
simply connected universal cover, $\tilde G= SU(n)$,  the dual group
may be defined in terms of the original group as follows.
We must use a metric to identify $\liet \cong \liet^*$.
For $SU(n)$, with the metric \metlie\  we have:
\eqn\dualgpii{
\Lambda_{coweight} (G^v) \equiv \Lambda_{weight} (G)
= \bigl[ \Lambda_{coweight}(G) \bigr] ^*
}
Specifically, $SU(nm)/\IZ_n$ is dual to $SU(nm)/\IZ_m$.

Let us now consider the predictions of $S$-duality. Invariance under
transformations of the type $\tau \rightarrow \tau+1$ {\it i.e.}
$\theta \rightarrow \theta+2\pi$ in \tgt\ are obviously symmetries.
The other generator of $S$-duality $\tau \rightarrow -1/\tau$
acts less trivially. $S$-duality
predicts that the theory defined as the Weyl-group orbifold of the
free field theory \xiii\ with the identifications \xiv\
for the group $G$ is equivalent to the theory
with $G$ replaced by its dual $ G^v$ and $\tau$ replaced by
$-1/\tau$. We now show that this is
identical to $T$-duality of the conformal field theory in \xiii.

First, let us
recall the conventions for $T$-duality.
Suppose $\Lambda, \Lambda^* \subset \IR^d$ are
dual lattices, where $\IR^d$ has the Euclidean
metric. Then standard $T$-duality states that the
theory
\eqn\cftiii{
\eqalign{
I^{1} & = {1\over 4 \pi \alpha} \int d^2 \sigma
\biggl[   \p_0 \vec X \cdot \p_0 \vec X
-
\p_1 \vec X \cdot \p_1 \vec X
\biggr] \cr
\vec X & \sim \vec X + 2 \pi \vec v \qquad \vec v\in \Lambda\cr}
}
is equivalent to the theory
\eqn\cftiiv{
\eqalign{
I^{2} & = {\alpha\over 4 \pi} \int d^2 \sigma
\biggl[  \p_0 \vec X \cdot \p_0 \vec X
-
\p_1 \vec X \cdot \p_1 \vec X
\biggr] \cr
\vec X & \sim \vec X + 2 \pi \vec v \qquad \vec v\in \Lambda^* \cr}
}
Since
 the  $G$ and $G^v$
theories reduce to
the  supersymmetric orbifolds based on the lattices:
\eqn\compare{
\eqalign{
G-{\rm theory}:\qquad &
 \Lambda_{coweight}(G)
\oplus  \Lambda_{coweight}(G) \cr
G^v-{\rm theory}:\qquad &
 \Lambda_{coweight}(G^v)
\oplus  \Lambda_{coweight}(G^v) \cr}
}
it is now manifest that  $S$-duality at $\rho_1=\theta=0,\rho_2=1$
 follows from \dualgpii. For other
values of $\rho,\theta$ the ``quadratic form'' defining the
action of the Gaussian model is given by
the matrix $E=B+G$ with
\foot{In
{\it Minkowskian} signature there is
no relative factor of $i$ between the kinetic
and topological terms.}
\eqn\xxiii{
E= \Biggl[\delta^{(6)}_{IJ} \oplus
q(\tau,\rho)
\Biggr] \otimes \Tr
}
where $q(\tau,\rho)$ is the $2\times 2$ matrix:
\eqn\qdfrm{
\eqalign{
q(\tau,\rho) &=
 {1\over \alpha} \sqrt{h}h_{ij} + {\theta\over 2 \pi} \epsilon_{ij}\cr
& = \pmatrix{{\tau_2\over \rho_2} \mid \rho\mid^2 & - \tau_2{\rho_1\over
\rho_2} + \tau_1\cr
- \tau_2{\rho_1\over \rho_2} - \tau_1 & {\tau_2\over \rho_2} \cr} \cr}
}
and $h$ is the two-metric in \llml.
Thus, the torus of the sigma model
has complexified K\"ahler form and
complex structure given by
$\tau$ and $\rho$, respectively.
Since
\eqn\qdfrmi{
q(-1/\tau,\rho) = \pmatrix{ 0 & -1\cr 1& 0\cr}
q(\tau,\rho)^{-1} \pmatrix{ 0 & 1\cr -1& 0\cr}
}
we see that the $S$-duality transformation
$\tau\to -1/\tau$ follows from the $T$-duality
transformation $E\to E^{-1}$ together with  the rotation
\eqn\rotation{
\pmatrix{\vec X^7\cr \vec X^8\cr} \rightarrow
\pmatrix{0 & -1\cr 1& 0\cr} \pmatrix{\vec X^7\cr \vec X^8\cr} \qquad .
}

It is useful to consider the simple example $G=SU(2)$ in more detail.
We can identify the coweight lattice of $SU(2)$ with
the root lattice $\Lambda_{root}$ of $su(2)$, which
we will take to be $\sqrt{2}$ times the integers.
The weight lattice of $SU(2)$,
$\Lambda_{weight}(SU(2))$ is then the integers divided
by $\sqrt{2}$ , and
$[ \Lambda_{weight}]^*= \Lambda_{root}$.
The dual of $SU(2)$ is $ G^v=SO(3)$. After
reduction the $SU(2)$  theory
contains
``momentum states''
created by the vertex operator
\eqn\estt{\cos\{{n\over \sqrt{2}} (X^8_L+X^8_R)\}
}
where $X^8_{L,R}$ are the left and right moving parts of $X^8$.
The time derivative of $X^8$ acting on such a state is non-zero.
 From the four-dimensional point of view,
since $X^8=A_3$ this means that
there is electric flux winding around the $x^3$ direction of the internal
torus. Under $T$-duality this is mapped to the winding state
\eqn\mstt{\cos\{ {n \over \sqrt{2}  } (X^8_L-X^8_R)\}
}
Now one finds that the spatial derivative
(in the $\sigma^1$ direction) of $X^8$ is nonzero. Thus there is a
magnetic flux in the $x^2$ direction. Note that this is
not quite in accord with $S$-duality. To recover
$S$-duality on the states we must compose
$T$-duality with the $\IZ_2$ transformation
\rotation, in accord with \qdfrmi.

If the worldsheet is a torus,  $\Sigma_1=T^2$
(or, more generally, has $\pi_1\not=0$) and
$X^7, X^8$ are in winding
number sectors:
\eqn\wndg{
\eqalign{
\vec X^7 & = 2 \pi \sigma^0 \vec v_0 + 2 \pi \sigma^1 \vec v_1\cr
\vec X^8 & = 2 \pi \sigma^0 \vec w_0 + 2 \pi \sigma^1 \vec w_1\cr}
}
with $v,w\in \Lambda_{coweight}(G)$
and $0< \sigma^0, \sigma^1 <1$, then the
$\theta$-dependent part of the
action is
\eqn\wndgi{
\theta \bigl[ (v_0, w_1) - (v_1, w_0) \bigr]
}
where we use the metric on $\liet$ in \metlie.
Thus, one can accordingly map winding number sectors
to instanton number sectors. Field
configurations with windings defining nontrivial
elements of  $\Lambda_{weight}/\Lambda_{root}$
for both $X^7,X^8$ satisfy 't Hooft-type boundary
conditions.

We now explore an
interesting parallel structure between 4D gauge
theories and 2D conformal field theory.
\foot{The connection between 4D supersymmetric
Yang-Mills theory and  conformal field theory
described in this paper is probably unrelated to the
connection  uncovered in \vw.}
A beautiful and famous phenomenon
in conformal field theory is the Frenkel-Kac construction, i.e.,
the existence of enhanced current algebra symmetries in  special
Gaussian models.  Our results suggest a 4D analog.  It is natural to conjecture
that
$N=4$ Yang-Mills in $D=4$ has enhanced symmetries
when the $SL(2,\IZ)$-action on the coupling is not free,
i.e., at $\tau= i,  e^{\pi i/3}$.
At these points the theory is strongly coupled and
$\theta=0,\pi$.
For gauge group $G=SU(2)$, the
conformal field theory we have described has enhanced
Kac-Moody symmetries of $SU(2) \times SU(2)$ or $SU(3)$ at these
points, before dividing by the Weyl group. A surviving
$SO(2)$ Kac-Moody symmetry in  the orbifold theory
 has a simple 4D interpretation. At
$\tau= i,  e^{\pi i/3}$ the theory is self-dual
and the gauge bosons are {\it degenerate} with spin one
monopoles. The isotropy group of the classical
$SL(2,\IR)$ symmetry acting on $\tau$ is  the classical
$SO(2)$ electric-magnetic rotation. Apparently,
this {\it continuous}
symmetry survives in the quantum theory at
$\tau= i,  e^{\pi i/3}$.
As for the other currents projected out of the orbifold theory
we may remark that,  in general, if a theory $A$ can be embedded in
a theory $B$ with symmetries it can happen that
the symmetries of $B$ strongly constrain the amplitudes of
$A$. Perhaps the $D=4$ theory has such hidden
symmetries at the strong-coupling points
$\tau= i,  e^{\pi i/3}$. Clearly, this
is an interesting topic for further work.

There are several other lines of investigation worth pursuing.
The infrared limit of our dimensional reduction  should be
studied more  carefully.
Generalizations to other  2-manifolds besides
the torus are of interest. Much stronger tests of  $S$-duality might
be obtained by considering the $L^2$ corrections to our
leading result. Even more ambitiously, perhaps a perturbative proof of
$S$-duality might be achieved  by analyzing the expansion to all orders.
Finally, the generalization of these remarks to theories
with $N=2$ (or fewer) supersymmetries promises to be
fascinating.

\bigskip
{\it Note added}: We would like to draw the
reader's attention to reference
\ref\bsv{M. Bershadsky, A. Johansen, V. Sadov and C. Vafa ,
``Topological Reduction of 4D SYM to 2D $\sigma$--Models,''
hep-th/9501096;  L. Girardello, A. Giveon, M. Porrati, and A.
 Zaffaroni, ``S-Duality in N=4 Yang-Mills Theories with General
Gauge Groups'', hep-th/9502057.  }
where issues similar to the above are
discussed.

\bigskip
\centerline{\bf Acknowledgements}

GM would like to thank E. Verlinde for some
initial collaboration on these matters.   We thank
D. Freed, H. Garland,
D. Kutasov, E. Martinec, N. Nekrasov, and S. Shatashvili
for useful remarks.
This work
was supported by DOE grants DE-FG02-92ER40704,
DOE-91ER40618, by NSF
grant PHY 91-23780
and by a Presidential Young Investigator Award.

\appendix{A}{Generalization to arbitrary compact groups}

The generalization of the above discussion to the case of
$G$ an arbitrary compact  group is straightforward.
Locally,
$G$ can be written as the product
$A\times K$ where $A$ is a torus and
$K$ is semisimple. $S$-duality for the part
of the theory associated
with the abelian factors of $G$ is elementary,
so we focus on $K$.  Quotients by finite
subgroups living in different factors yield
orbifold versions of the conformal field theories derived below,
so, henceforth, we take $G$ to be connected
and simple.

\subsec{Normalization of the action}

We normalize the the action so that
anti-self-dual instantons in
the simply connected covering group $\tilde G$
associated to $G$ have action $2\pi ik \tau$, with
the instanton number $k$ taking on all integral
values.  This gives the normalization:
\eqn\genmet{
\Tr(ab) = \half B_\lieg(\upsilon,\upsilon) \Phi_\lieg(a,b)
}
where $\Phi_\lieg(\cdot, \cdot)$ is the Killing form on
$\lieg$, $B_\lieg(\cdot, \cdot)$ is the induced
form on $\lieg^*$ and $\upsilon$ is the highest
root of $\lieg$. \foot{In this paper $V^*$ indicates
the dual of a vector space, i.e., the space of
linear functionals on $V$. A quadratic form $Q$ on
$V$ canonically defines a form on $Q^*$ on $V^*$. If we
choose bases the two forms are inverse matrices.}

\subsec{Review of the dual group}

We briefly review the general definition
of the magnetic group $G^v$ dual to a
compact Lie group $G$
\refs{\gno,\olivrev}. We first
distinguish  two kinds of quantum numbers in a
gauge theory with unbroken gauge group $G$:
electric and magnetic. These are defined by
representation theory and topology, respectively,
as follows:

{\it Electric} quantum numbers are
given by representations of $G$. Representations are
determined by characters  $\chi$. By conjugation,
$\chi$ is completely
determined by its restriction to the maximal
torus $T\subset G$.  Thus,
the electric quantum numbers live on the
lattice $ \hat{T} \equiv Hom(T,U(1)) $.
Using the exponential map we may
think of this lattice as being in $\liet^*$:
$\Lambda_{weight}(G)\subset \liet^*$.

{\it Magnetic} quantum numbers are
related to $G$ bundles over
$S^2$. These are
determined by the equatorial transition function
$g(\phi):S^1\to G$.
By conjugation, $g(\phi)$  can be taken to
be in $T$.  Thus,
the magnetic quantum numbers live on the
lattice $ \check{T} \equiv Hom(U(1),T) $.
Using the exponential map we may
think of this lattice as being in $\liet$:
$\Lambda_{coweight}(G)\subset \liet$.

Notice that since $Hom(U(1),U(1))=\IZ$
the weight and coweight lattices of $G$,
hence electric and magnetic quantum numbers,
are {\it canonically} dual :
\eqn\crtiii{
\Lambda_{coweight}(G) = \bigl[\Lambda_{weight}(G) \bigr]^*
}
This is the Dirac quantization condition.
\foot{We adopt standard notation whereby, if
$V$ is a vector space,  the reciprocal
lattice $\Lambda^*\subset V^* $ to the lattice $\Lambda\subset V$
 is the lattice of vectors with integer pairings:
$
\Lambda^* = \{ v\in V^* : \forall w\in
\Lambda,  \quad \langle v,  w\rangle \in \IZ
\}
$ . }

Physically we may define the dual group as
follows.
Given any  compact Lie group $G$, the dual group
$G^v$ is the group for which the electric and
magnetic lattices are exchanged \gno. It is a
nontrivial fact that $G^v$ exists for every compact
group $G$.
Mathematically,
 the dual group is best understood
by thinking of a Lie algebra as defined by  its
root system $R$, following \bourbaki\humphreys.
We assume $\lieg$ is semisimple.
Let $V$ be a vector space.  A finite subset
$R\subset V$ is a root system if it satisfies
certain axioms \bourbaki\humphreys. One key axiom
states that  for all $\alpha\in R$
$\exists !$ $\alpha^v\in V^*$ with $\langle \alpha,\alpha^v\rangle
=2$.  The axioms are completely symmetric between
$R\subset V$ and the set $R^v\subset V^*$.
Now, to a root system $R$ (and a choice of simple roots) we
associate a Lie algebra $\lieg(R)$ defined by the Serre presentation.
Since root systems come in pairs
$R,R^v$ we get two dual Lie algebras $\lieg(R)$ and
$\lieg(R^v)$. \foot{The Cartan matrices are related by
transposition, except for $G_2$ where one must
reorder the simple roots.}
Furthermore, by construction, we have {\it canonically}:
\eqn\csas{
\eqalign{
V& = \liet^*(R) = \liet(R^v) \cr
V^* & = \liet (R) = \liet^*(R^v) \cr}
}
Finally, we can define the dual groups $G,G^v$.
These have simply connected covers
corresponding to $\lieg(R)$ and
$\lieg(R^v)$, respectively, and have global
structure such that
\eqn\dulgpiii{
\matrix{
\Lambda_{coweight}(G^v) & = & \Lambda_{coweight}(G)^*\cr
\cap &  & \cap \cr
 \liet(R^v)  & = &  \liet^*(R)\cr}
}
so the lattice of magnetic quantum numbers
of $G$ becomes the lattice of electric quantum
numbers of $G^v$, and vice versa.

\subsec{Checking S-duality}

The Gaussian model in conformal field theory
is defined by
a  triple of data $(V,Q,\Lambda)$, where
 $V$ is a vector space with nondegenerate
quadratic form $Q$ and
 $\Lambda\subset V$ is a lattice. The action is
\eqn\abgss{
\eqalign{
I & = {1\over 4 \pi} \int  d ^2\sigma
\biggl[Q( \p_0 \vec X, \p_0 \vec X)
-
Q( \p_1 \vec X ,  \p_1 \vec X)
\biggr] \cr
\vec X & \sim \vec X + 2 \pi \vec v \qquad \vec v\in \Lambda\cr}
}
$T$-duality is the equivalence of the
triples $(V,Q, \Lambda)$ and $(V^*, Q^* , \Lambda^*)$.

We can now show that $T$-duality implies
$S$-duality
restricted to the effective conformal field theory.
{}From \dulgpiii\ we see that the vector
spaces and lattices are naturally dual. The metric
associated to $G$ follows from \genmet. Moreover
for a dual pair of groups $G=G(R)$, $G^v=G(R^v)$
associated to root systems $R,R^v$
we have $\Phi_{\lieg(R)}^* = B_{\lieg(R^v)}$ and
a simple  identity \bourbaki\
\eqn\finalstp{
{B_{\lieg(R^v)}(\upsilon,\upsilon)\over 2 }
\Phi_{\lieg(R^v)} (\cdot, \cdot)
 = {2\over B_{\lieg(R)}(\upsilon,\upsilon)}
B_{\lieg(R^v) }(\cdot, \cdot)
}
shows that the metrics are inversely related.
Nonzero values of $\rho_1, \theta$ are handled
in the same way as in the $SU(n)$ case explained
above. Finally, $G$ and $G^v$ have
canonically isomorphic Weyl groups $W(G) = W(G^v)$,
so the $T$-dual orbifold of the dimensionally
reduced $G$-theory is
identical to the  dimensional reduction of
the $S$-dual $G^v$-theory.

\listrefs

\end